# Direct observation of giant binding energy modulation of exciton complexes in monolayer MoSe$_2$

Garima Gupta[1], Sangeeth Kallatt[1,2], and Kausik Majumdar[1*]

[1]Department of Electrical Communication Engineering, Indian Institute of Science, Bangalore 560012, India

[2]Center for Nano Science and Engineering, Indian Institute of Science, Bangalore 560012, India

[*]Corresponding author, email: kausikm@iisc.ac.in

**ABSTRACT:** Screening due to surrounding dielectric medium reshapes the electron-hole interaction potential and plays a pivotal role in deciding the binding energies of strongly bound exciton complexes in quantum confined monolayers of transition metal dichalcogenides (TMDs). However, owing to strong quasi-particle bandgap renormalization in such systems, a direct quantification of estimated shifts in binding energy in different dielectric media remains elusive using optical studies. In this work, by changing the dielectric environment, we show a conspicuous photoluminescence (PL) peak shift at low temperature for higher energy excitons (2s, 3s, 4s, 5s) in monolayer MoSe$_2$, while the 1s exciton peak position remains unaltered - a direct evidence of varying compensation between screening induced exciton binding energy modulation and quasi-particle bandgap renormalization. The estimated modulation of binding energy for the 1s exciton is found to be $58.6\%$ ($72.8\%$ for 2s, $75.85\%$ for 3s, $85.6\%$ for 4s) by coating an Al$_2$O$_3$ layer on top, while the corresponding reduction in quasi-particle bandgap is estimated to be 246 meV. Such a direct evidence of large tunability of the binding energy of exciton complexes as well as the bandgap in monolayer TMDs holds promise of novel device applications.

Monolayer TMDs exhibit strongly bound exciton complexes [1–7] even at room temperature due to enhanced electron-hole interaction resulting from strong out-of-plane quantum confinement [8], large in-plane carrier effective mass [9,10], and small dielectric constant [9,11]. To be able to externally control their binding energy will provide an unprecedented opportunity to design novel exciton based devices. Owing to the ultra-thin nature of the monolayer, a change in the surrounding dielectric medium is expected to provide a direct way to achieve this through screening of the electron-hole coulomb interaction [12–20]. However, despite such expected change in the binding energy of the exciton complexes, several studies [1,18,19,21–24] report that the PL emission peak position of the 1s exciton remains practically unaltered, irrespective of the surrounding dielectric environment. This apparently bemusing result is generally understood as a consequence of strong quasi-particle bandgap renormalization resulting from screened many body interaction [18,19,23]. The change in the electron-hole interaction strength of the 1s exciton bound state in presence of different dielectric media is almost equal to the change in the electron-electron interactions. Such compensation makes it challenging to experimentally validate the predicted effect of dielectric environment on the binding energy of excitons using optical studies in an unambiguous way [1,18–24]. While a few indirect approaches, including combinations of photoluminescence studies [18,19] with first principles calculations [17–19,25] and scanning tunneling spectroscopy [19,26,27], have been reported in the recent past, a direct evidence of binding energy modulation in monolayer TMDs by changing the surrounding dielectric media is still lacking. Also, a generalization of the compensation effect between bandgap renormalization and binding energy change for higher energy exciton states remains unclear.

In this work, by using low temperature PL spectra of monolayer MoSe$_2$ embedded in different dielectric environments, we observe that unlike the unchanged PL emission peak position of 1s exciton state, the higher energy (2s, 3s, 4s, 5s) exciton states show monotonically increasing peak shifts in response to changed surroundings. Further, the charged trion binding energy is also found to exhibit a strong shift. The experimental observations are explained by a screening model with spatially distributed charge. The results confirm that a change in surrounding dielectric medium induces a strong quasiparticle bandgap modification in conjunction with a strong binding energy modulation of different exciton complexes. While the 1s exciton binding energy change is almost exactly equal in magnitude to the quasiparticle bandgap change, the higher energy exciton states exhibit larger percentage change in binding energy.

The origin of reduction in binding energy in presence of surrounding medium of higher dielectric constant can be conceptually understood as a consequence of the proximity of an image charge, as schematically illustrated in Figure 1a. The additional repulsive force due to the image charge effectively weakens the strong in-plane electron-hole coulomb attraction, in turn reducing the binding energy of various excitonic bound states. Monolayer flakes of MoSe$_2$ are mechanically exfoliated on clean Si substrate, covered with 300 nm SiO$_2$. In some of the samples, 10 nm thick Al$_2$O$_3$ film was deposited by using electron beam evaporation at a chamber pressure of $2\times10^{-6}$ mBar. Figures 1b shows the acquired PL spectra of MoSe$_2$/SiO$_2$ (sample M1) and Al$_2$O$_3$/MoSe$_2$/SiO$_2$ (sample M2) stacks at $T = 8$ K. The 1s state of neutral $A$ exciton ($A^0_{1s}$) and the corresponding charged trion ($A^-_{1s}$) peaks (only negatively charged trions are considered due to slight n-type doping of the sample) are found to be at similar energies in both the samples. To confirm this observation further, we measure the PL emission of monolayer MoSe$_2$ at room temperature, embedded in five different dielectric surroundings. The PL spectra showing the $A^0_{1s}$ peak is presented in Figure 1c, confirming a negligible peak shift ($\approx 10$ meV). While the enhanced electrostatic screening due to surrounding dielectric reduces the exciton binding energy ($\Delta E_b$), at the same time, this screens and modifies the electron-electron interaction term in the many-body Hamiltonian as well. This results in a reduction ($\Delta E_g$) in the quasi-particle bandgap. The non-tunability of the $A^0_{1s}$ peak for different dielectric environments suggests that the change in energy of this state due to these two effects are equal and opposite, almost entirely compensating for each other for the 1s exciton, as schematically illustrated in the left panel of Figure 1d. However, on moving to states with increasing principal quantum number ($n$), $\Delta E_b$ reduces gradually, whereas $\Delta E_g$ remains the same, and hence such exact compensation is not expected. This possibility allows for a direct observation of PL peak shift for these higher energy states with a change in dielectric environment, as explained in the middle and right panel of Figure 1d.

Information regarding the higher energy states is obtained by careful observation of the PL spectrum of sample M1 at different temperatures, as shown in Figure 2a-d. We identify four peaks by fitting the experimental data in the energy range of 1.75 eV to 2.0 eV. The small peak (in purple) around 30 meV below the prominent $B_{1s}^0$ peak (in green) is attributed to the $B_{1s}^-$ trion peak. Note that, the $A_{1s}^-$ trion peak intensity is much stronger than the $A_{1s}^0$ exciton peak, while the $B_{1s}^-$ trion is significantly weaker than the $B_{1s}^0$ exciton peak. This difference can be explained from the origin of lowest energy bright and dark trion states, as schematically illustrated in Figure 2e-f. The availability of the second electron in the spin split lower conduction band required to form $B_{1s}^-$ trion is bottlenecked as most of these electrons are consumed to form the more favorable *A* exciton and trion. In addition, the dipole allowed bright trion $B_{1s,B}^-$, that contributes to the PL signal, is at higher energy compared to the dark trion $B_{1s,D}^-$ and hence forms with lower probability. On the other hand, for the A trions, their energy positions get interchanged, making $A_{1s,B}^-$ bright trion a more favorable trion state than the dark trion $A_{1s,D}^-$. This also explains the observation of higher intensity of $B_{1s}^-$ peak at higher temperature as shown in Figures 2a-d as the fractional contribution of bright trions increases with temperature.

The conspicuous peak around 1.8 eV is assigned to $A_{2s}^0$ exciton. 2s exciton peak at similar energy has been reported for monolayer MoSe$_2$ in recent reflectance experiment [28]. We also observe the existence of another higher energy peak due to $A_{3s}^0$ exciton around 1.9 eV.

In the bottom panel of Figure 3a, the peak positions of $A_{1s}^0$ through $A_{5s}^0$, and $B_{1s}^0$ and $B_{1s}^-$ states for sample M2 are shown. The overall peak intensities in this sample are weaker than sample M1, but are clearly distinguishable. From the peak positions, we estimate a strong red shift of 50.3 meV and 104.1 meV in the $A_{2s}^0$ and $A_{3s}^0$ peaks respectively due to addition of a top Al$_2$O$_3$ coating at $T = 8$ K. The higher energy peaks become difficult to distinguish beyond $T = 35$ K. Figure 3b shows a comparison between the $A_{1s}^0$ PL emission peak position for both samples, measured at different temperatures and suggests that the non-tunability of its energy is temperature independent. In Figure 3c, the temperature dependent $A_{2s}^0$ and $A_{3s}^0$ transition energies confirm strong red shift of PL peaks in sample M2 at various temperatures. The Bohr radius of these higher energy states is significantly larger compared to 1s exciton. Consequently, the magnitude of change in binding energy becomes increasingly weak with higher $n$ in presence of Al$_2$O$_3$ coating. This leaves the bandgap renormalization effect partially uncompensated, explaining the red shift (Figure 1d).

To have a quantitative understanding, we compute the effective potential due to a hole by solving Poisson's equation for the five-layer dielectric structure schematically shown in Figure 4a. The solution of Poisson's equation in the $x - z$ plane automatically takes care of the image force effects arising from differences in dielectric constants among different layers. The dielectric constant and the exciton effective mass $\mu_{ex}$ is taken as 4.74 and 0.31m$_0$ [9]. To account for the finite spread of the carrier wave function, the point charge is replaced by a Gaussian distribution in the plane $(x, y)$ of the monolayer, modulated by the square of the wave function of the first eigen state in a square quantum well in the out of plane ($z$) direction, and is given by:

$$\rho(x,y,z) = \frac{q}{\sqrt{2\pi\sigma^2}} exp\left(\frac{-(x^2+y^2)}{2\sigma^2}\right) \cos^2\left(\frac{\pi z}{t}\right)$$

The thickness ($t$) of the monolayer is assumed to be 6.5Å. Solution of Poisson's equation provides the converged potential profile $V$ for this distributed hole charge [29]. The energy eigenvalues $E$ and wave functions $\psi(x,y)$ of the exciton bound states are obtained by numerically solving the two-dimensional time independent Schrodinger equation in the $x-y$ plane [30]:

$$\left(-\frac{\hbar^2}{2\mu_{ex}}\left(\frac{\partial^2}{\partial x^2}+\frac{\partial^2}{\partial y^2}\right) - qV(x,y)\right)\psi(x,y) = E\psi(x,y)$$

The in-plane spread $\sigma$ is used as a single fitting parameter, and $\sigma = 22.5$Å (both for samples M1 and M2) provides the best fitting with the experimental data in Figure 5. The corresponding charge density and the in-plane potential profiles are shown in Figure 4b-c. See Supplemental Material [31] for the binding energies and shape of the wave functions of first few bound states.

We estimate the quasiparticle bandgap energy (continuum) for both the stacks as: $E_g = A_{1s}^0(experiment) + E_b^{1s}(model)$. The calculated PL emission energies are plotted as a function of $n$ in Figure 5 for both samples, and are in good agreement with the experimental peak positions. The difference between the computed continuum levels in the samples immediately leads to an estimation of $\Delta E_g \approx 246$ meV due to quasi-particle bandgap renormalization. The estimated reduction in the $A_{1s}^0$ binding energy due to addition of top $Al_2O_3$ coating is 58.6%. The corresponding reductions for $A_{2s}^0$ and $A_{3s}^0$ are 72.8% and 75.85%, respectively. The measured PL peak position and estimated binding energy of different states are tabulated in Table I.

Table I: Binding energy of different excitonic states at $T = 8$ K in sample M1 and M2

| Type of Exciton | Binding Energy (meV) MoSe$_2$/SiO$_2$ | Al$_2$O$_3$/MoSe$_2$/SiO$_2$ | Change in Binding Energy |
|---|---|---|---|
| $A^0_{1s}$ | 420 ($\pm 0.37$) | 174 | 58.6% |
| $A^0_{2s}$ | 276 ($\pm 0.93$) | 75 ($\pm 0.021$) | 72.8% |
| $A^0_{3s}$ | 188 ($\pm 0.02$) | 45.4 ($\pm 0.038$) | 75.85% |
| $A^0_{4s}$ | 117* | 16.8 ($\pm 0.025$) | 85.64% |
| $A^0_{5s}$ | 74* | 10.3 ($\pm 0.053$) | 86% |
| $A^-_{1s}$ | 32.3 | 27.4 | 15.1% |

* Model predicted

In Table I, we have also shown the measured $A^-_{1s}$ trion binding energy: $\Delta^-_{1s} = A^0_{1s} - A^-_{1s}$, which also exhibits 15.1% reduction in sample M2. Note that, $\Delta^-_{1s}$ being estimated from the separation between two PL peaks, does not involve the quasi-particle bandgap, and provides an independent direct evidence of the modulation of binding energy in presence of larger dielectric screening.

In conclusion, we explored new perspective of environment screening on two-dimensional monolayers by exploiting the idea of increasing mismatch between quasiparticle bandgap renormalization and modification in exciton binding energy for increasing quantum number of the exciton state. The proposed technique allows us to unambiguously estimate all the necessary information about the excitonic series and quasi-particle bandgap change in two-dimensional monolayer embedded in different dielectric media. Our results clearly demonstrate the prominent effect of substrate and environment induced screening in two-dimensional system, making this effect crucial to be taken into account while analyzing results in existing devices based on 2D materials. For example, the band structure of the 2D material in the region underneath the contact material or in the presence of a gate dielectric is expected to be modified locally as a result of this effect, and is expected to play an important role in determining the device performance. Similar revisit will also be required in analyzing the performance of 2D material based photodetectors as this screening induced unintentionally created built-in field at the source junction will support efficient electron-hole separation. Finally, planar hetero-junctions in two-dimensional crystals are generally difficult to achieve due to stringent growth conditions. The results described in this work open up the possibility of a new class of two-dimensional planar heterostructure devices by only spatially modifying the substrate or top dielectric constant.

ACKNOWLEDGMENT

The authors acknowledge the help from S. Nair and B. Prahlada in sample fabrication and characterization. The authors acknowledge the support of nano-fabrication and characterization facilities at CeNSE, IISc. K.M. would like to acknowledge the support of a start-up grant from IISc, Bangalore; the support of a grant under Indian Space Research Organization (ISRO); grants under Ramanujan fellowship, Early Career Award, and Nano Mission under Department of Science and Technology (DST), Government of India; and a young faculty grant from Robert Bosch Center for Cyber Physical System.

**References:**

[1] Z. Ye, T. Cao, K. O'Brien, H. Zhu, X. Yin, Y. Wang, S. G. Louie, and X. Zhang, Nature **513**, 214 (2014).

[2] K. He, N. Kumar, L. Zhao, Z. Wang, K. F. Mak, H. Zhao, and J. Shan, Phys. Rev. Lett. **113**, 1 (2014).

[3] J. S. Ross, S. Wu, H. Yu, N. J. Ghimire, A. M. Jones, G. Aivazian, J. Yan, D. G. Mandrus, D. Xiao, W. Yao, and X. Xu, Nat. Commun. **4**, 1474 (2013).

[4] S. Kallatt, G. Umesh, and K. Majumdar, J. Phys. Chem. Lett. **7**, 2032 (2016).

[5] K. F. Mak, K. He, C. Lee, G. H. Lee, J. Hone, T. F. Heinz, and J. Shan, Nat. Mater. **12**, 207 (2013).

[6] Y. You, X.-X. Zhang, T. C. Berkelbach, M. S. Hybertsen, D. R. Reichman, and T. F. Heinz, Nat. Phys. **11**, 477 (2015).

[7] X. Zhang, T. Cao, Z. Lu, Y. Lin, F. Zhang, Y. Wang, Z. Li, J. C. Hone, J. A. Robinson, D. Smirnov, and S. G. Louie, **2**, (n.d.).

[8] K. F. Mak, C. Lee, J. Hone, J. Shan, and T. F. Heinz, Phys. Rev. Lett. **105**, 2 (2010).

[9] A. Ramasubramaniam, Phys. Rev. B - Condens. Matter Mater. Phys. **86**, 1 (2012).

[10] K. Majumdar, C. Hobbs, and P. D. Kirsch, IEEE Electron Device Lett. **35**, 402 (2014).

[11] H. P. Komsa and A. V. Krasheninnikov, Phys. Rev. B - Condens. Matter Mater. Phys. **86**, 1 (2012).

[12] A. Chernikov, T. C. Berkelbach, H. M. Hill, A. Rigosi, Y. Li, O. B. Aslan, D. R.

Reichman, M. S. Hybertsen, and T. F. Heinz, Phys. Rev. Lett. **113**, 1 (2014).

[13] L. V. Keldysh, Sov. Phys. JETP **29**, 658 (1979).

[14] M. Kleefstra and G. C. Herman, J. Appl. Phys. **51**, 4923 (1980).

[15] M. Rösner, C. Steinke, M. Lorke, C. Gies, F. Jahnke, and T. O. Wehling, Nano Lett. **16**, 2322 (2016).

[16] X. Zhu, N. R. Monahan, Z. Gong, H. Zhu, K. W. Williams, and C. A. Nelson, J. Am. Chem. Soc. **137**, 8313 (2015).

[17] I. Kylänpää and H. P. Komsa, Phys. Rev. B - Condens. Matter Mater. Phys. **92**, 1 (2015).

[18] Y. Lin, X. Ling, L. Yu, S. Huang, A. L. Hsu, Y. H. Lee, J. Kong, M. S. Dresselhaus, and T. Palacios, Nano Lett. **14**, 5569 (2014).

[19] M. M. Ugeda, A. J. Bradley, S.-F. Shi, F. H. da Jornada, Y. Zhang, D. Y. Qiu, W. Ruan, S.-K. Mo, Z. Hussain, Z.-X. Shen, F. Wang, S. G. Louie, and M. F. Crommie, Nat. Mater. **13**, 1091 (2014).

[20] A. Raja, A. Chaves, J. Yu, G. Arefe, M. Hill, A. F. Rigosi, T. C. Berkelbach, P. Nagler, C. Schüller, T. Korn, C. Nuckolls, J. Hone, and E. Louis, arXiv:1702.01204v2 1 (2017).

[21] M. Buscema, G. A. Steele, H. S. J. Van Der Zant, and A. Castellanos-gomez, **7**, 561 (2014).

[22] Y. Li, Z. Qi, M. Liu, Y. Wang, X. Cheng, G. Zhang, and L. Sheng, Nanoscale **6**, 15248 (2014).

[23] B. Liu, W. Zhao, Z. Ding, I. Verzhbitskiy, L. Li, J. Lu, J. Chen, G. Eda, and K. P. Loh, Adv. Mater. 6457 (2016).

[24]   A. V. Stier, N. P. Wilson, G. Clark, X. Xu, and S. A. Crooker, Nano Lett. **16**, 7054 (2016).

[25]   T. C. Berkelbach, M. S. Hybertsen, and D. R. Reichman, Phys. Rev. B - Condens. Matter Mater. Phys. **88**, 1 (2013).

[26]   C. Zhang, Y. Chen, A. Johnson, M. Y. Li, L. J. Li, P. C. Mende, R. M. Feenstra, and C. K. Shih, Nano Lett. **15**, 6494 (2015).

[27]   W. Zhang, Q. Wang, Y. Chen, Z. Lin, A. Mccreary, and N. Briggs, 2D Mater. **4**, 15021 (2017).

[28]   A. Arora, K. Nogajewski, M. Molas, M. Koperski, and M. Potemski, Nanoscale **7**, 20769 (2015).

[29]   H. Ehrenreich and M. H. Cohen, Phys. Rev. **115**, 786 (1959).

[30]   X. L. Yang, S. H. Guo, F. T. Chan, K. W. Wong, and W. Y. Ching, Phys. Rev. A **43**, 1186 (1991).

[31]   See Supplemental Material at the end for the calculated binding energies and the corresponding shape of the wave functions of the first few bound states, both for $MoSe_2/SiO_2$ and $Al_2O_3/MoSe_2/SiO_2$ stacks.

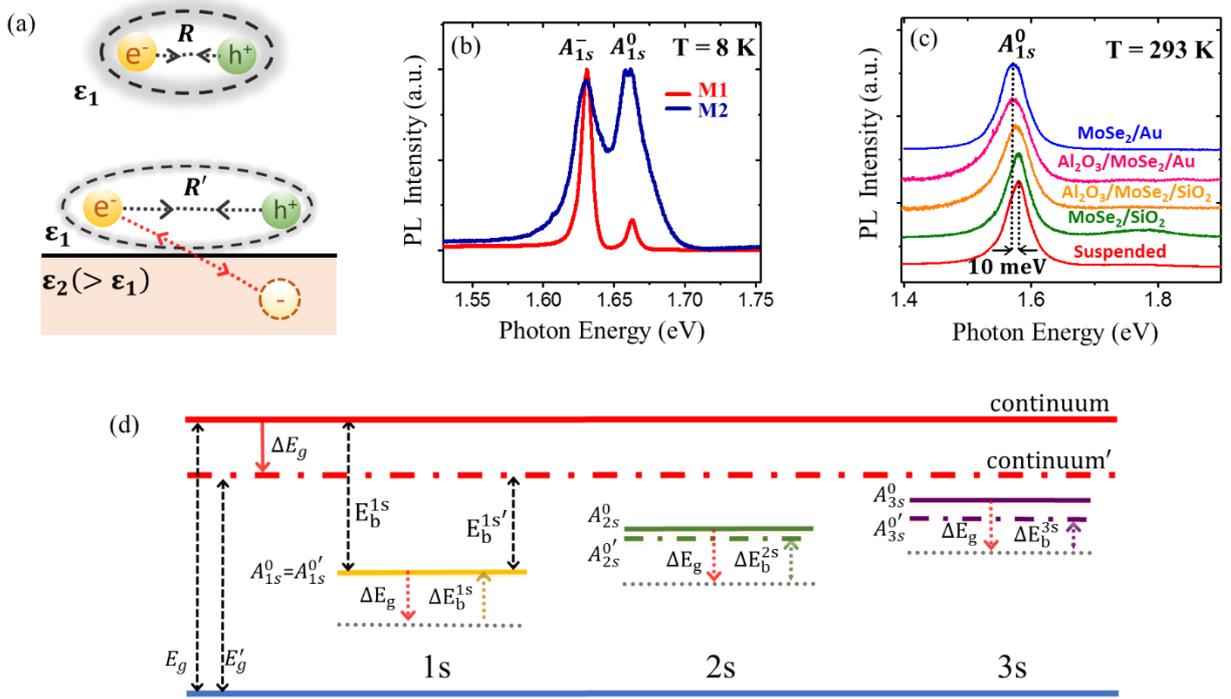

**Figure 1.** (a) Schematic representation of the effect of image charge arising from introduction of a dielectric layer in the vicinity of monolayer MoSe$_2$, with an enhancement of exciton Bohr radius. (b) PL spectra of single layer MoSe$_2$, acquired at 8 K, for samples M1 and M2. $A_{1s}^0$ and $A_{1s}^-$ represent the neutral 1s A exciton and corresponding negatively charged trion. (c) PL spectra of monolayer MoSe$_2$ obtained at room temperature for different top and bottom dielectric materials. The $A_{1s}^0$ exciton peak energy is nearly unaffected with screening, showing a maximum of 10 meV shift. (d) An increase in screening modifies the quasi-particle bandgap ($\Delta E_g$), bringing the continuum down to continuum$'$, and reduces the binding energy of exciton states. These two effects compete and decide the final transition energy of various exciton states upon modified surroundings. The compensation is nearly equal for 1s exciton state, resulting in unchanged $A_{1s}^0$ energy $\left(\Delta E_g = \Delta E_b^{1s}\right)$. The binding energy reduction decreases in magnitude for higher energy exciton states $\left(|\Delta E_b^{1s}| > |\Delta E_b^{2s}| > |\Delta E_b^{3s}|\right)$, resulting in only partial compensation of the renormalized bandgap induced shift. This suggests observation of red shift in $A_{2s}^0$ and $A_{3s}^0$ peaks when surrounded by medium with large dielectric constant.

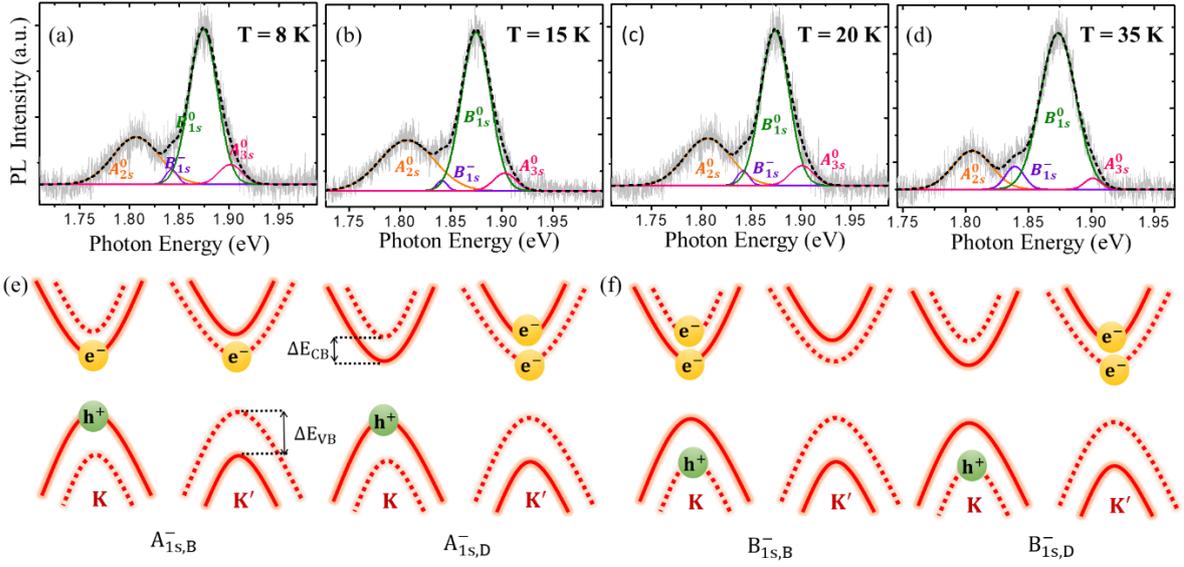

**Figure 2.** (a)-(d) The experimental PL data for sample M1 (in grey) in the energy range 1.75 eV to 2.0 eV, measured at sample temperatures ranging from 8 K to 35 K. Four peaks are obtained from fitting the data, namely $A_{2s}^0$ exciton (orange), $A_{3s}^0$ exciton (pink), $B_{1s}^-$ trion (purple), $B_{1s}^0$ exciton (green). The $B_{1s}^-$ trion peak, about 30 meV below $B_{1s}^0$, shows slight increase in intensity with temperature. (e) Schematic of lowest energy bright and dark A trion states. The dotted and solid lines in conduction band (valence band) correspond to spin up (down) and spin down (up) states of electron (hole) in monolayer MoSe$_2$. The bright state $A_{1s,B}^-$ is at lower energy than the $A_{1s,D}^-$ dark state. (f) The lowest energy bright $\left(B_{1s,B}^-\right)$ and dark $\left(B_{1s,D}^-\right)$ B trion suggests $B_{1s,B}^-$ state at higher energy than the dark $B_{1s,D}^-$ trion state, which also explains the slight increase in $B_{1s,B}^-$ intensity with temperature in (a)-(d).

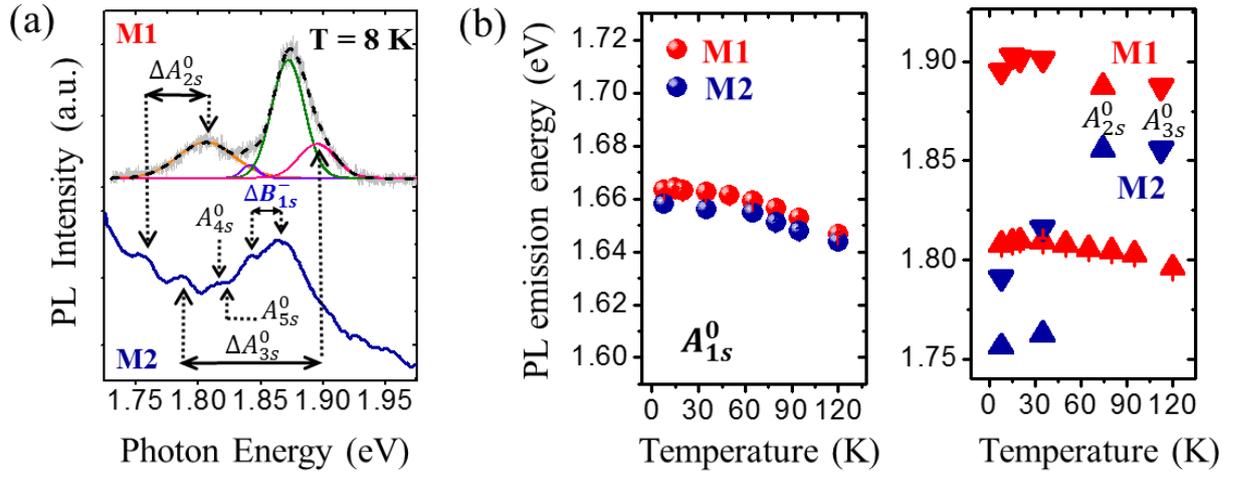

**Figure 3.** (a) PL peak positions for higher energy A exciton states ($A_{2s}^0$ through $A_{5s}^0$) in sample M2. Clear red shift for $A_{2s}^0$ and $A_{3s}^0$ peaks are observed in M2 compared with M1. $B_{1s}^-$ trion peak close to $B_{1s}^0$ peak for both samples is also discernable. (b) The $A_{1s}^0$ exciton peak is almost similar for samples M1 and M2 at various measurement temperatures. (c) The $A_{2s}^0$, $A_{3s}^0$ PL peak positions for M1 and M2 at different temperatures testify strong PL peak shift for higher energy exciton states.

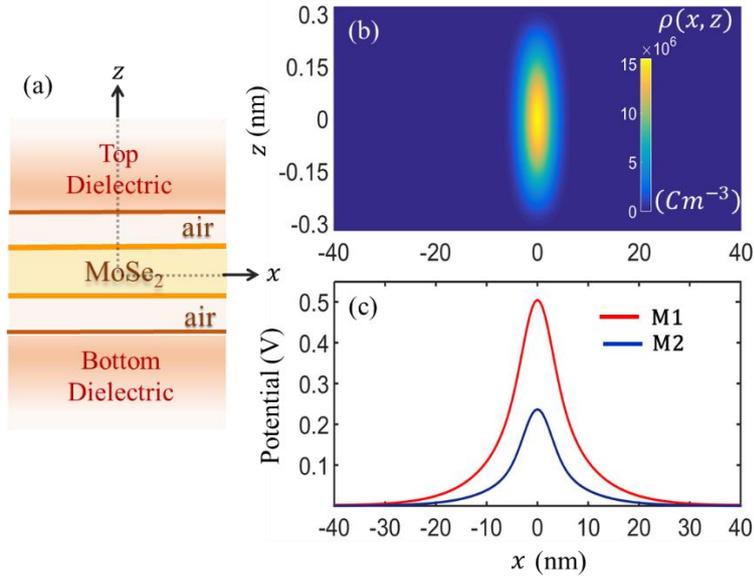

**Figure 4.** (a) Schematic of five-dielectric-layer configuration considered for quantitative binding energy estimations of excitons in ML MoSe$_2$ including 3Å van der Waals gap. (b) The hole distribution in the $x - z$ cross-section plane of MoSe$_2$. (c) Converged potential due to the distributed hole in the plane of the monolayer MoSe$_2$ for sample M1 and M2.

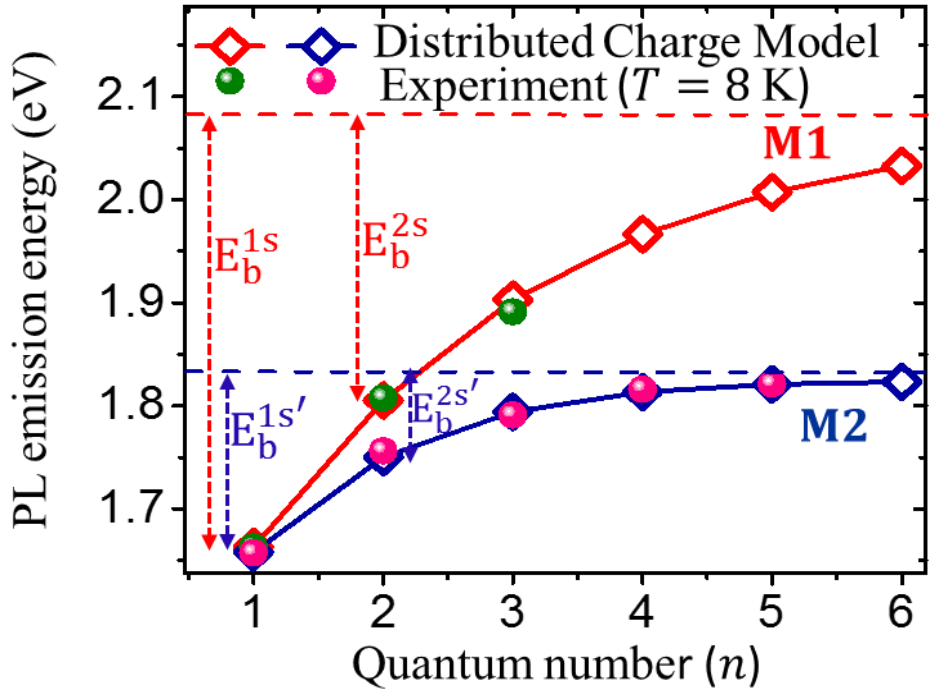

**Figure 5.** Experimental PL peak positions (solid circles) and the corresponding values predicted from the distributed charge model (open symbol). The continuum levels are represented by dashed lines (red for sample M1 and blue for sample M2). The binding energy of different exciton states is extracted from the difference between the continuum and the PL peak position, as schematically illustrated for the 1s and 2s states.

# Supplemental Material:

# Direct observation of giant binding energy modulation of exciton complexes in monolayer MoSe$_2$

Garima Gupta[1], Sangeeth Kallatt[1,2], and Kausik Majumdar[1*]

[1]Department of Electrical Communication Engineering, Indian Institute of Science, Bangalore 560012, India

[2]Center for Nano Science and Engineering, Indian Institute of Science, Bangalore 560012, India

[*]Corresponding author, email: kausikm@iisc.ac.in

Calculated binding energy values in eV for excitons with various principal ($n$) and orbital angular momentum ($l$) quantum number for (a) MoSe$_2$/SiO$_2$ and (b) Al$_2$O$_3$/MoSe$_2$/SiO$_2$ stacks:

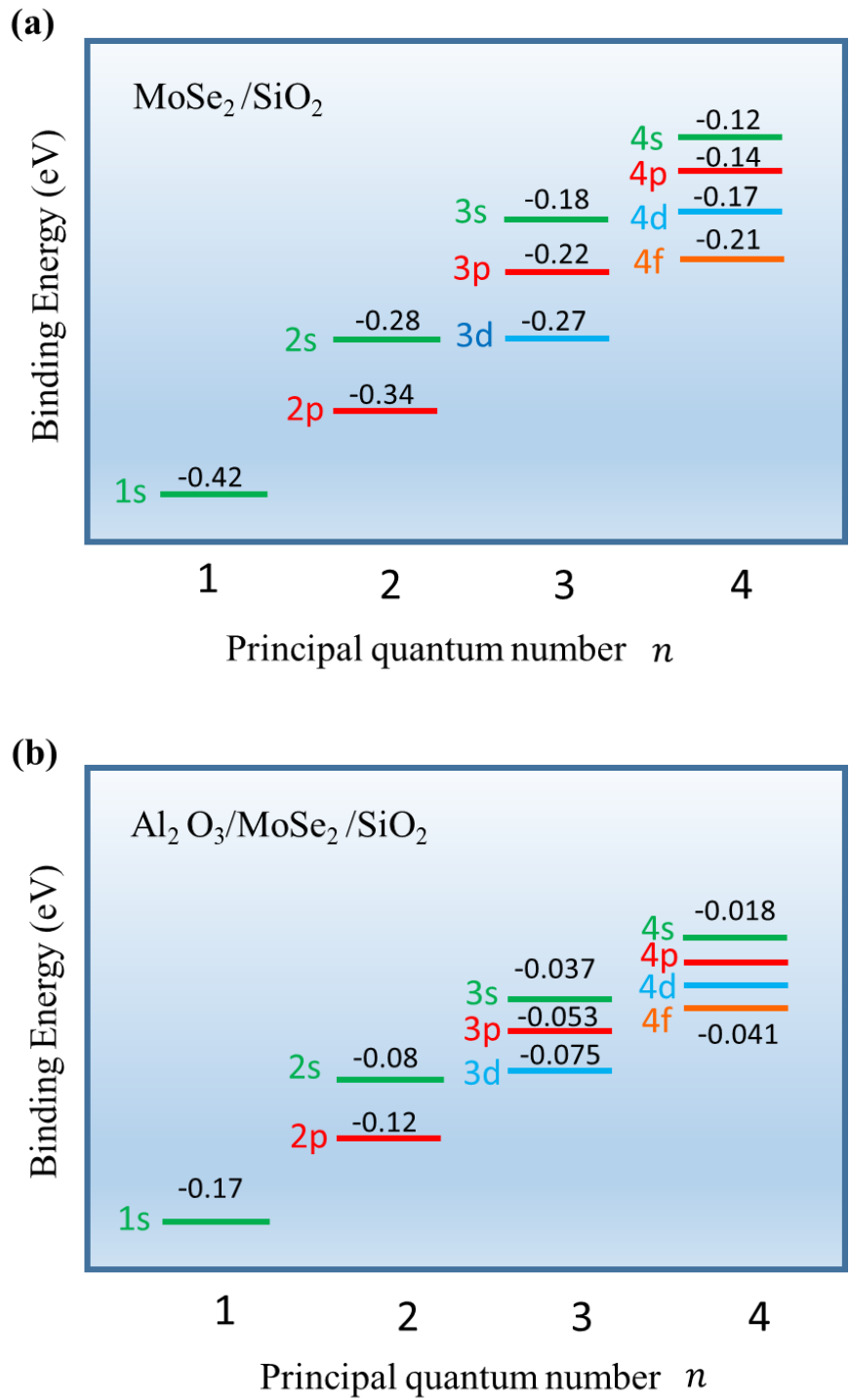

Wave functions of the first few bound eigenstates

● Positive ● Negative  ● Positive ● Negative

| MoSe$_2$/SiO$_2$ | $n = 1$ | Al$_2$O$_3$/MoSe$_2$/SiO$_2$ |
|---|---|---|
| 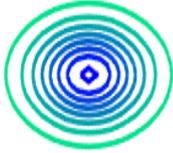 | $l = 0$ | 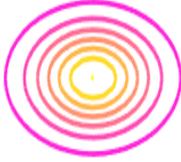 |
| | $n = 2$ | |
| 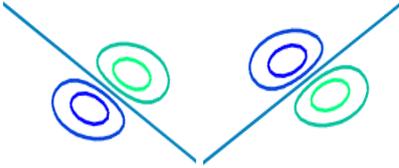 | $l = \pm 1$ | 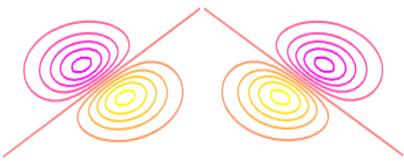 |
| 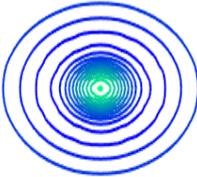 | $l = 0$ | 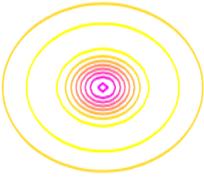 |
| | $n = 3$ | |
| 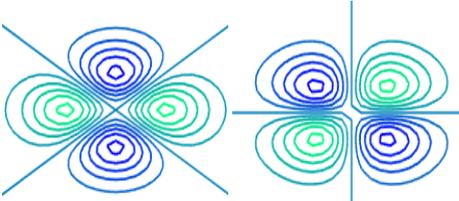 | $l = \pm 2$ | 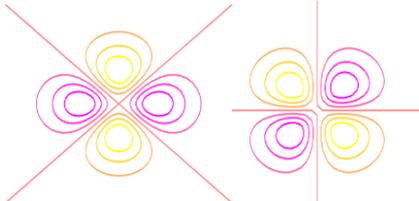 |
| 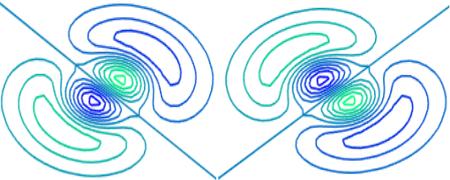 | $l = \pm 1$ | 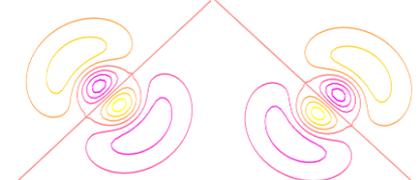 |
| | $n = 4$ | |
| 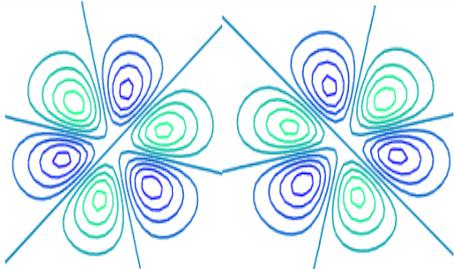 | $l = \pm 3$ | 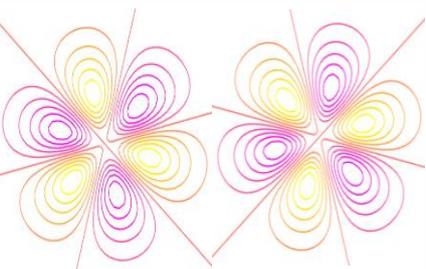 |